\documentclass[aps,preprint,pra,showpacs]{revtex4}
\usepackage{graphicx}
\usepackage[pdftex]{color}

\begin{document}

\title{Stronger-than-quantum bipartite correlations violate relativistic causality in the classical limit}

\author{Daniel Rohrlich}
\address{Department of Physics, Ben Gurion University of the Negev, Beersheba
84105 Israel}

\date{\today}

\begin{abstract}
Superquantum (``PR-box'') correlations, though designed to respect relativistic causality, $violate$ relativistic causality in the classical limit.  Generalizing to all stronger-than-quantum bipartite correlations, I derive Tsirelson's bound from the axioms of nonlocality, relativistic causality and the existence of a classical limit.  This derivation of Tsirelson's bound does not assume quantum mechanics yet suggests how Hilbert space is implicit in quantum correlations.
\end{abstract}

\pacs{03.65.Ta, 03.30.+p, 03.65.Ud, 03.67.Hk}

\maketitle

The logical structure of the special theory of relativity is exemplary:  two axioms, each with a clear physical meaning, are so nearly incompatible that a unique kinematics reconciles them. In comparison, the axioms of quantum mechanics \cite{von} are opaque.  Aharonov \cite{a} suggested, by analogy with special relativity, that also quantum mechanics might follow from two axioms with clear physical meanings:  nonlocality and relativistic causality.  Quantum nonlocality comprises nonlocal equations of motion \cite{nleq} and nonlocal correlations; here we focus on correlations.  Quantum correlations are nonlocal---they violate the Bell-CHSH \cite{bell} inequality---but respect relativistic causality---they do not transmit superluminal signals.  Nonlocality and relativistic causality seem incompatible, yet quantum mechanics reconciles them.  Is quantum mechanics $unique$ in reconciling them (as Shimony \cite{s} independently suggested)?  Can we derive quantum mechanics from these two axioms?   Popescu and Rohrlich \cite{PR} answered this question in the negative by defining hypothetical ``superquantum" correlations that (unlike quantum correlations) violate the Bell-CHSH inequality maximally, while respecting relativistic causality.  Others \cite{o,ic} have shown that the axioms of nonlocality and relativistic causality, together with an additional axiom (or a stronger axiom of relativistic causality called ``information causality" \cite{ic}), rule out superquantum (or ``PR-box") correlations, and come close to ruling out all stronger-than-quantum correlations.  However, the physical meaning of these axioms is obscure.  By contrast, I claim that relativistic causality, nonlocality, and a minimal additional axiom with clear physical meaning---namely, the existence of a classical limit---together rule out all stronger-than-quantum correlations (and not just PR-box correlations as in Ref. \cite{prncl}).  Navascu{\'e}s and Wunderlich \cite{nw} have published a similar claim, but define the classical limit quite differently, via the ``wiring" \cite{bs} of entangled systems, and not via incompatible measurements that become compatible in the classical limit.

How is the axiom of a classical limit minimal?  We ask whether it is possible to generalize quantum mechanics while respecting relativistic causality, or whether quantum mechanics is unique.  More specifically, we ask whether nonlocal correlations could violate the Bell-CHSH inequality more strongly than the quantum bound, Tsirelson's bound \cite{ts}, while respecting relativistic causality.  But quantum mechanics has a classical limit.  In this limit there are no noncommuting quantum observables; there are only jointly measurable macroscopic observables.  This classical limit---our direct experience---is an inherent constraint, a kind of boundary condition, on quantum mechanics and on any generalization of quantum mechanics.  Thus stronger-than-quantum correlations, too, must have a classical limit.

Consider a setting for measuring nonlocal correlations:  Alice and Bob share pairs of particles on which they measure observables $a$, $a^\prime$, $b$ and $b^\prime$.  One particle in each such pair is in Alice's lab, and she measures either $a$ or $a^\prime$ (but not both); the other particle is in Bob's lab, and he measures either $b$ or $b^\prime$ (but not both).  The result of each measurement is $\pm 1$, and Alice and Bob measure at spacelike separations.  After measurements on many pairs, they pool their data and discover PR-box correlations \cite{PR}:
\begin{equation}
C(a,b)=C(a,b^\prime)=C(a^\prime, b) =1=-C(a^\prime, b^\prime)~~~,
\label{e1}
\end{equation}
where $C(a,b)$ is the correlation between Alice's measurements of $a$ and Bob's measurements
of $b$, etc.  By definition,
\begin{equation}
C(a,b) = {\rm{p}}_{ab}(1,1) +{\rm{p}}_{ab}(-1,-1) - {\rm{p}}_{ab}(1,-1) -
{\rm{p}}_{ab}(-1,1)~~~,
\label{e2}
\end{equation}
where ${\rm{p}}_{ab}(i,j)$ is the probability that measurements of $a$ and $b$ yield $a=i$ and
$b=j$.  PR-box correlations violate the Bell-CHSH inequality
\begin{equation}
\left\vert C(a,b) +C(a,b^\prime) +C(a^\prime,b) - C(a^\prime,b^\prime) \right\vert\le 2
\label{BCHSH}
\end{equation}
maximally, since $C(a,b)+ C(a,b^\prime) +C(a^\prime, b)-C(a^\prime, b^\prime)=4$.  In addition, Alice and Bob each discover that their respective measurements of $a$, $a^\prime$ and $b$, $b^\prime$ are equally likely to yield $\pm 1$ regardless of what the other measures.  It follows that Alice cannot send a signal to Bob by her choice of what to measure on a pair, and likewise Bob cannot send a signal to Alice.  Hence PR-box correlations violate the Bell-CHSH inequality maximally while respecting relativistic causality.  But now consider the classical limit.

First, note that if Alice measures $a$ and obtains 1, she can predict with certainty that Bob will obtain 1 whether he measures $b$ or $b^\prime$; if she obtains $-1$,  she can predict with certainty that he will obtain $-1$ whether he measures $b$ or $b^\prime$.  (By contrast, quantum correlations would allow Alice to predict with certainty only the result of measuring $b$ or the result of measuring $b^\prime$ but not both \cite{EPRB}.)  If Alice measures $a^\prime$, she can predict with certainty that Bob will obtain her result if he measures $b$ and the opposite result if he
measures $b^\prime$.  Thus, all that protects relativistic causality is the (assumed) complementarity between $b$ and $b^\prime$:  Bob cannot measure both, although---from Alice's point of view---no uncertainty principle governs $b$ and $b^\prime$.

Next, suppose that Alice measures $a$ or $a^\prime$ on $N$ pairs.  Let us define macroscopic observables $B$ and $B^\prime$:
\begin{equation}
B={{b_1+ b_2 +\dots+b_N}\over N}~~~~,~~~B^\prime={{b^\prime_1+ b^\prime_2 +\dots+
b^\prime_N}\over N}~~~,
\label{defbbp}
\end{equation}
where $b_m$ and $b^\prime_m$ represent $b$ and $b^\prime$, respectively, on the $m$-th pair. Alice already knows the values of $B$ and $B^\prime$, and there must be ``weak" measurements (analogous to weak measurements in quantum mechanics \cite{weak}) that Bob can make to obtain partial information about $both$ $B$ and $B^\prime$; for, in the classical limit, there can be no complementarity between $B$ and $B^\prime$.  Now it is true that $a=1$ and $a=-1$ are equally likely, and so the average values of $B$ and $B^\prime$ vanish, whether Alice measures $a$ or $a^\prime$.  But if she measures $a$ on each pair, then typical values of $B$ and $B^\prime$ will be $\pm 1/\sqrt{N}$ (but possibly as large as $\pm 1$) and correlated.  If she measures $a^\prime$ on each pair, then typical values of $B$ and $B^\prime$ will be $\pm 1/\sqrt{N}$ (but possibly as large as $\pm 1$) and
$anti$-correlated.  Thus Alice can signal a single bit to Bob by consistently choosing whether to measure $a$ or $a^\prime$.  This claim is delicate because the large-$N$ limit in which $B$ and $B^\prime$ commute is also the limit that suppresses the fluctuations of $B$ and $B^\prime$.  To ensure that Bob has a good chance of measuring $B$ and $B^\prime$ accurately enough to determine whether they are correlated or anti-correlated, $N$ may have to be large and therefore the fluctuations in $B$ and $B^\prime$ will be small.  However, Alice and Bob can repeat this experiment (on $N$ pairs at a time) as many times as it takes to give Bob a good chance of catching and measuring large enough fluctuations.  They can repeat the experiment exponentially many times (exponentially in $N$).  Alice and Bob's expenses and exertions are not our concern. Relativistic causality does not forbid superluminal signalling only when it is cheap and reliable.  Relativistic causality forbids superluminal signalling altogether.

Since PR-box correlations were defined without a classical limit, we cannot specify exactly how the approach to the classical limit depends on $N$.  But this is no objection.  What matters is only that when Bob detects a correlation, it is more likely that Alice measured $a$ than when he detects an anti-correlation.  If it were not more likely, it would mean that Bob's measurements yield zero information about $B$ or about $B^\prime$, contradicting the fact that there is a classical limit in which $B$ and $B^\prime$ are jointly measurable.
  
For example, let us suppose Bob considers only those sets of $N$ pairs in which $B=\pm1$ and $B^\prime=\pm 1$.  The probability of $B=1$ is $2^{-N}$.  But if Alice is measuring $a$ consistently, the probability of $B=1$ $and$ $B^\prime =1$ is also $2^{-N}$, and not $2^{-2N}$, while the probability of $B=1$ and $B^\prime =-1$ vanishes.  If Alice is measuring $a^\prime$ consistently, the probabilities are reversed.  (These probabilities must be folded with the scatter in Bob's measurements, but the scatter is independent of what Alice measures.)  Thus with unlimited resources, Alice can send a (superluminal) signal to Bob. Superquantum (PR-box) correlations are $not$ consistent with relativistic causality in the classical limit.

It is not just PR-box correlations that violate relativistic causality in the classical limit.  Suppose that instead of Eq.\ (\ref{e1}) we have
\begin{equation}
C(a,b)=C(a,b^\prime)=C(a^\prime, b) =C=-C(a^\prime, b^\prime)~~~,
\label{e3}
\end{equation}
where $-1\le C\le 1$.  If $C$ is close enough to 1, it will still be possible for Alice to send Bob superluminal signals in the classical limit.  However, we cannot make this claim as $C$ decreases.  There will be some critical value of $C$ at which the correlations in Eq.\ (\ref{e3}) become compatible with relativistic causality.  We would like to know if the critical value coincides with the quantum value, $C_Q ={\sqrt{2}}/2$.  Can we calculate this critical $C$?

In answering this question, we begin with a calculation that contains an (instructive) error.  First, we reformulate the result for PR boxes \cite{prncl} as follows.  Suppose that Bob measures $B+B^\prime$ on each set of $N$ pairs (to some precision).  No matter what Alice measures, Bob's results must average out to $\langle B +B^\prime\rangle=0$.  But if Alice measures $a^\prime$ on all the pairs, then $B+B^\prime =0$ identically for each set of $N$ pairs.  If Alice measures $a$ on all the pairs, then the values of $B+B^\prime$ on successive sets of $N$ pairs fall in a binomial distribution centered at 0.  The distribution that Bob measures is a convolution of the distribution Alice generates and the inherent scatter in his measurements of $B+ B^\prime$.  As long as the scatter is not a flat distribution---and it is not flat in the classical limit---Bob will be able to detect whether Alice is measuring $a$ or $a^\prime$ and she will be able to send him a superluminal signal.  Therefore PR-box correlations violate relativistic causality in the classical limit.

Next, consider correlations of Eq.\ (\ref{e3}) with $C$ close to 1.  If Alice measure $a^\prime$ consistently, she generates a distribution of $B+B^\prime$ that is no longer identically zero, but still not as wide as the distribution of $B+B^\prime$ she generates if she measures $a$ consistently, and the distributions Bob measures in the two cases will differ.  Reducing $C$, however, widens the distribution of the anticorrelated results and narrows the distribution of the correlated results.  At some critical value of $C$, Alice will not be able to send Bob any superluminal signal.  For $C<1$, we cannot derive the precise distribution of $B+B^\prime$; but we can derive limits on how wide the distribution of $B+B^\prime$ could be.  That is, we can infer the range of the variance $[\Delta (B+B^\prime)]^2$ of $B+B^\prime$.  Since the average value $\langle B+B^\prime\rangle$ vanishes, the variance is $\langle ( B+B^\prime )^2 \rangle$.  To calculate $\langle ( B+B^\prime )^2 \rangle$ precisely, we need the probabilities of the results $-2$, 0 and 2 for $b+b^\prime$.  Can we obtain these probabilities from the correlation $C$?  

Suppose Alice measures $a$ on a given pair.  From $C(a,b)=C$ we know that the probability of $a=b$ is $p_+=(1+C)/2$, and the probability of $a=-b$ is $p_-=(1-C)/2$.  (These values follow from $p_++p_-=1$ and $p_+ -p_-=C$.)  Likewise, from $C(a,b^\prime)=C$ we know that the probability of $a=b^\prime$ is $p_+=(1+C)/2$, and the probability of $a=-b^\prime$ is $p_-=(1-C)/2$.  But we still don't know the probabilities of the results $-2$, 0 and 2 for $b+b^\prime$.  It could be that $a=b$ and $a=b^\prime$ always coincide, and $a=-b$ and $a=-b^\prime$ always coincide; then $b$ and $b^\prime$ would remain perfectly correlated (unlike $a$ and $b$, $b^\prime$) implying $\Delta (B+B^\prime) = \Delta (b+b^\prime)=2$.  The opposite limit would be for $b$ and $b^\prime$ to differ as often as possible.  Even so, $b$ and $b^\prime$ cannot differ with probability greater than $2p_-=1-C$, as Fig. \ref{Fig1}(a-b) shows; hence the minimum probability of $b+b^\prime = \pm 2$ is $C$.  Letting $\Delta_a (B+B^\prime)$ denote the standard deviation in $B+B^\prime$ when Alice measures $a$ consistently, we have $\Delta_a (B+B^\prime) \ge 2\sqrt{C}$.

Analogously, if Alice measures $a^\prime$ on a given pair, we know from $C(a^\prime ,b)=C$ that the probability of $a^\prime=b$ is $p_+=(1+C)/2$, and the probability of $a^\prime=-b$ is $p_-=(1-C)/2$.  Likewise, from $C(a^\prime,b^\prime)=-C$ we know that the probability of $a^\prime=-b^\prime$ is $p_+=(1+C)/2$, and that the probability of $a^\prime=b^\prime$ is $p_-=(1-C)/2$.  But we don't know the probabilities of the results $-2$, 0 and 2 for $b+b^\prime$.  It could be that $a^\prime=\pm b$ and $a^\prime=\mp b^{\prime}$ always coincide, in which case $b$ and $b^\prime$ remain perfectly anticorrelated (unlike $a^\prime$ and $b^\prime$) implying $\Delta (B+B^\prime) = \Delta (b+b^\prime)=0$.  The opposite limit would be for $b$ and $b^\prime$ to coincide as often as possible.  Even so, $b$ and $b^\prime$ cannot coincide with probability greater than $2p_-=1-C$, as Fig. \ref{Fig1}(c-d) shows; hence the maximum probability of $b+b^\prime = \pm 2$ is $1-C$.  Letting $\Delta_{a^\prime} (B+B^\prime)$ denote the standard deviation in $B+B^\prime$ when Alice measures $a^\prime$ consistently, we have $\Delta_{a^\prime} (B+B^\prime) \le 2\sqrt{1-C}$.

In general, the variances that Bob obtains from his measurements of $B+B^\prime$ exceed the ranges generated by Alice; but as long as $\Delta_a (B+B^\prime) \ne \Delta_{a^\prime} (B+B^\prime)$ and their resources are unlimited, Alice can still send Bob a superluminal signal.  Thus, to eliminate the possibility of superluminal signalling, a necessary condition is $\Delta_a (B+B^\prime) = \Delta_a^\prime (B+B^\prime)$; and the maximal value $C$ that satisfies this condition is $C=1-C$, i.e. $C= 1/2$.  But this value is $below$ the quantum value $C_Q$ and satisfies the Bell-CHSH inequality!

The error in this calculation is that we have tacitly assumed that $b$ and $b^\prime$ add as scalars.  If they add as scalars, then indeed $b+b^\prime = 0$ or $\pm 2$.  But now let us drop this assumption, and consider $B+B^\prime$ and $B-B^\prime$ without making assumptions about $b+b^\prime$ and $b-b^\prime$. All the same, since $B$ and $B^\prime$ exist in the classical limit, so do their sum and difference.

As before, we calculate variances according to the formula $(\Delta X)^2 = \langle X^2\rangle -\langle X\rangle^2$, where the angle brackets $\langle ~{\hbox{  }} \rangle$ denote statistical averaging.  Instead of assuming Eq.\ (\ref{e3}), let us consider completely general correlations $C(a,b)$, $C(a,b^\prime)$, $C(a^\prime,b)$, $C(a^\prime, b^\prime)$.  Measurements of $a$ and $a^\prime$ are still equally likely to yield $\pm 1$, however; so the averages $\langle B\rangle$, $\langle B^\prime \rangle$ and therefore also $\langle B\pm B^\prime \rangle$ vanish, whatever Alice measures.  We cannot calculate $\Delta_{a^\prime} (B+ B^\prime)$ directly, but we note that
\begin{equation}
\langle (B+ B^\prime)^2\rangle + \langle (B- B^\prime)^2\rangle = 2 \langle B^2\rangle
+2\langle (B^\prime)^2\rangle~~~,
\label{rot}
\end{equation}
regardless of what Alice measures, and the calculation of $\langle B^2\rangle$ and $\langle (B^\prime)^2\rangle$ is straightforward:
\begin{equation}
\langle B^2\rangle  = {{\langle b_1^2\rangle +\dots+\langle b_N^2 \rangle}\over {N^2}} =
{1\over N} ={{\langle (b^\prime_1)^2\rangle +\dots+\langle
(b^\prime_N)^2 \rangle}\over {N^2}} =\langle (B^\prime)^2\rangle~~~,
\label{bbp}
\end{equation}
where we have neglected cross-terms $\langle b_j b_k\rangle$ for $j\ne k$ because $b_j$ and
$b_k$ are uncorrelated, and likewise $\langle b^\prime_j b^\prime_k\rangle$.  We have
\begin{equation}
\left[ \Delta_{a^\prime} (B + B^\prime)\right]^2 +\left[ \Delta_{a^\prime} (B - B^\prime)\right]^2 = 2\langle B^2\rangle +2\langle (B^\prime)^2\rangle = 4/N~~~~.
\label{trans}
\end{equation}
Therefore we can replace $\left[ \Delta_{a^\prime} (B+B^\prime)\right]^2$ with $4/N-\left[ \Delta_{a^\prime} (B-B^\prime)\right]^2$.  Now relativistic causality requires $\Delta_a (B+ B^\prime)=\Delta_{a^\prime} (B+ B^\prime)$, thus $\left[ \Delta_a (B + B^\prime)\right]^2 = {4/ N}-\left[ \Delta_{a^\prime} (B - B^\prime)\right]^2$, i.e.
\begin{equation}
\left[ \Delta_a (B + B^\prime)\right]^2 + \left[ \Delta_{a^\prime} (B-B^\prime)\right]^2 =4/N~~~~.
\label{quad}
\end{equation}

Back to Alice and Bob.  Suppose Alice measures $a$ on many groups of $N$ pairs. She obtains $a=\pm 1$ with equal probability in each measurement.  Bob measures $B+B^\prime$ on each group of $N$ pairs.  The average result is $\langle B+B^\prime\rangle =0$, because results for $B$ and $B^\prime$ are distributed symmetrically around 0.  But we want to calculate also the variance in $B +B^\prime$.  We will calculate it (as well as we can) by defining $A=(a_1 +a_2+\dots+a_N)/N$ and comparing the distribution of $A$ as measured by Alice with the distribution of $B+B^\prime$ as measured by Bob.  The distribution of $A$ is a simple binomial:  the possible values of $A$ are precisely $A= -1, -(N-2)/N, \dots , (N-2)/N, 1$, with probability $N! /2^Nn! (N-n)!$ for a given $A = 1-2n/N$.  An essential difference between the distributions of $A$ and of $B+B^\prime$ is that while the values of $A$ are restricted to precisely $N+1$ values, the values of $B+ B^\prime$ are not.  For example, if Alice obtains $A=1$ (i.e. she obtains $a=1$ for all $N$ pairs), then Bob obtains $B+B^\prime \approx C(a,b)+C(a,b^\prime)$.  But $C(a,b)$ and $C(a, b^\prime)$ are defined in the limit of infinitely many pairs; for finite $N$, we cannot assume that $B+B^\prime$ is precisely $C(a,b)+C(a, b^\prime)$.  In general, when Alice obtains $A=1- 2n/N$, Bob obtains $B+B^\prime \approx (1-2n/N)\left[ C(a,b) +C(a,b^\prime) \right]$.

So, on the one hand, we cannot $assume$ that the values $B+B^\prime$ are precisely \hbox{$(1-2n /N)$} $\left[ C(a,b) +C(a,b^\prime) \right]$.  On the other hand, it is a logical possibility that $A$ is more strongly correlated with $B+B^\prime$ than with either $B$ or $B^\prime$, and even that $A$ and $B+B^\prime$ are perfectly correlated.  If $A$ and $B+B^\prime$ are perfectly correlated, then we are justified in assigning to each of the pairs an observable $c$ (on Bob's end) taking values $\pm \left[C(a,b)+C(a, b^\prime)\right]$, and assuming that it is perfectly correlated with the observable $a$ for that pair.  In this case, the variance in Bob's measurements of $B+B^\prime$ is simply the variance of this binomial distribution:  $[\Delta_a (B+B^\prime)]^2=\left[C(a,b)+C(a, b^\prime)\right]^2/{N}$.  The other logical possibility is that $A$ and $B+B^\prime$ are $not$ perfectly correlated.  In this case, all we can say about the variance is that it is greater:  $[\Delta_a (B+B^\prime)]^2>\left[C(a,b)+ C(a, b^\prime)\right]^2/{N}$.  It cannot be equal to or less than $\left[C(a,b)+C(a, b^\prime) \right]^2/{N}$ because it includes the binomial variance in Alice's measurements of $A$ plus additional variance relative to Alice's results.  To summarize, we can write
\begin{equation}
\left[C(a,b)+C(a, b^\prime)\right]/\sqrt{N} ~\le~ \Delta_a (B+B^\prime)~~~,
\label{db}
\end{equation}
with equality $only$ in the case that $A$ and $B+B^\prime$ are perfectly correlated and we can say that $a_m$ and $c_m$ are perfectly correlated for all $m$.  The calculation of $\Delta_{a^\prime} (B - B^\prime)$ is completely analogous.  (Note that it does not matter whether Bob actually measures $B - B^\prime$ or not.)  The bound corresponding to Eq.\ (\ref{db}) is
\begin{equation}
\left[C(a^\prime,b)-C(a^\prime, b^\prime)\right]/\sqrt{N} ~\le~ \Delta_{a^\prime} (B-B^\prime)~~~,
\label{dbp}
\end{equation}
with equality $only$ in the case that $A^\prime=(a^\prime_1+a^\prime_2+\dots+a^\prime_N)/N$ and $B-B^\prime$ are perfectly correlated and we can say that $a^\prime_m$ and some $c_m^\prime$ are perfectly correlated for all $m$.

Combining Eqs.\ (\ref{quad}-\ref{dbp}), we write
\begin{equation}
\left[C(a,b)+C(a, b^\prime)\right]^2+\left[C(a^\prime,b)-C(a^\prime, b^\prime)\right]^2 \le 4
\label{condi}
\end{equation}
as a requirement of relativistic causality.  Without loss of generality, we can assume that $C(a,b) +C(a, b^\prime)$ and $C(a^\prime,b) - C(a^\prime,b^\prime)$ have the same sign.  (If not, we could simply interchange $b$ and $b^\prime$ throughout.)  Then, applying \cite{kb} the inequality $|x+y|\le ({2x^2+2y^2})^{1/2}$ with $x= C(a,b)+C(a, b^\prime)$ and $y=C(a^\prime, b)-C(a^\prime, b^\prime)$, we obtain Tsirelson's bound:
\begin{equation}
\left\vert C(a,b) +C(a,b^\prime) +C(a^\prime,b) - C(a^\prime,b^\prime) \right\vert \le 2\sqrt{2}~~~~.
\label{kerndi}
\end{equation}
Quantum mechanics saturates this bound with $C(a,b) =C(a,b^\prime) =C(a^\prime,b) =C_Q= - C(a^\prime,b^\prime)$, where $C_Q= \sqrt{2}/2$.  We have derived a theorem of quantum mechanics---Tsirelson's bound---directly from the axioms of nonlocality, relativistic causality and the existence of a classical limit.

Indeed, we have obtained more than Tsirelson's bound.  We that if $b$ and $b^\prime$ add as scalars, then their sum $b+ b^\prime$ can be only 0 or $\pm 2$, and the no-signalling constraint implies local correlations.  But if the linear combinations $c/ \vert C(a,b)+ C(a,b^\prime) \vert=(b+b^\prime) /\vert C(a,b)+C(a,b^\prime) \vert$ and $c^\prime/ \vert C ( a^\prime , b)-C(a^\prime ,b^\prime)\vert= (b- b^\prime )/\vert C(a^\prime ,b)- C (a^\prime ,b^\prime)\vert$ of observables $b, b^\prime$ taking values $\pm 1$ are themselves observables taking values $\pm 1$---namely, if they add as components of vectors---then the resulting correlations are nonlocal and even saturate Tsirelson's bound.  Thus, quantum correlations derive from a Hilbert space:  if the observables $b$ and $b^\prime$ add as components of vectors, the values of $c$ and $c^\prime$ can be $\pm \vert C(a,b)+C(a,b^\prime) \vert$ and $\pm \vert C(a^\prime ,b)-C(a^\prime , b^\prime) \vert$, respectively, rather than 2, 0 or $-2$.  Thus, one day our answer to the question, ``Why is quantum mechanics so weird?" may be, ``Quantum mechanics is the way it is because it is the most nonlocal theory consistent with relativistic causality in the classical limit."  It will be a more satisfying answer than von Neumann's list of opaque axioms \cite{von}.

\begin{acknowledgments}

I thank Yakir Aharonov for deep insights that have informed this work.  I thank Yemima Ben-Menahem and Miguel Navascu{\'e}s for stimulating correspondence and Sandu Popescu, Nathan Argaman and Nicolas Gisin for critical comments.

\end{acknowledgments}

\begin{figure}
\end{figure}

\begin{figure}
\centerline{
\includegraphics*[width=120mm]{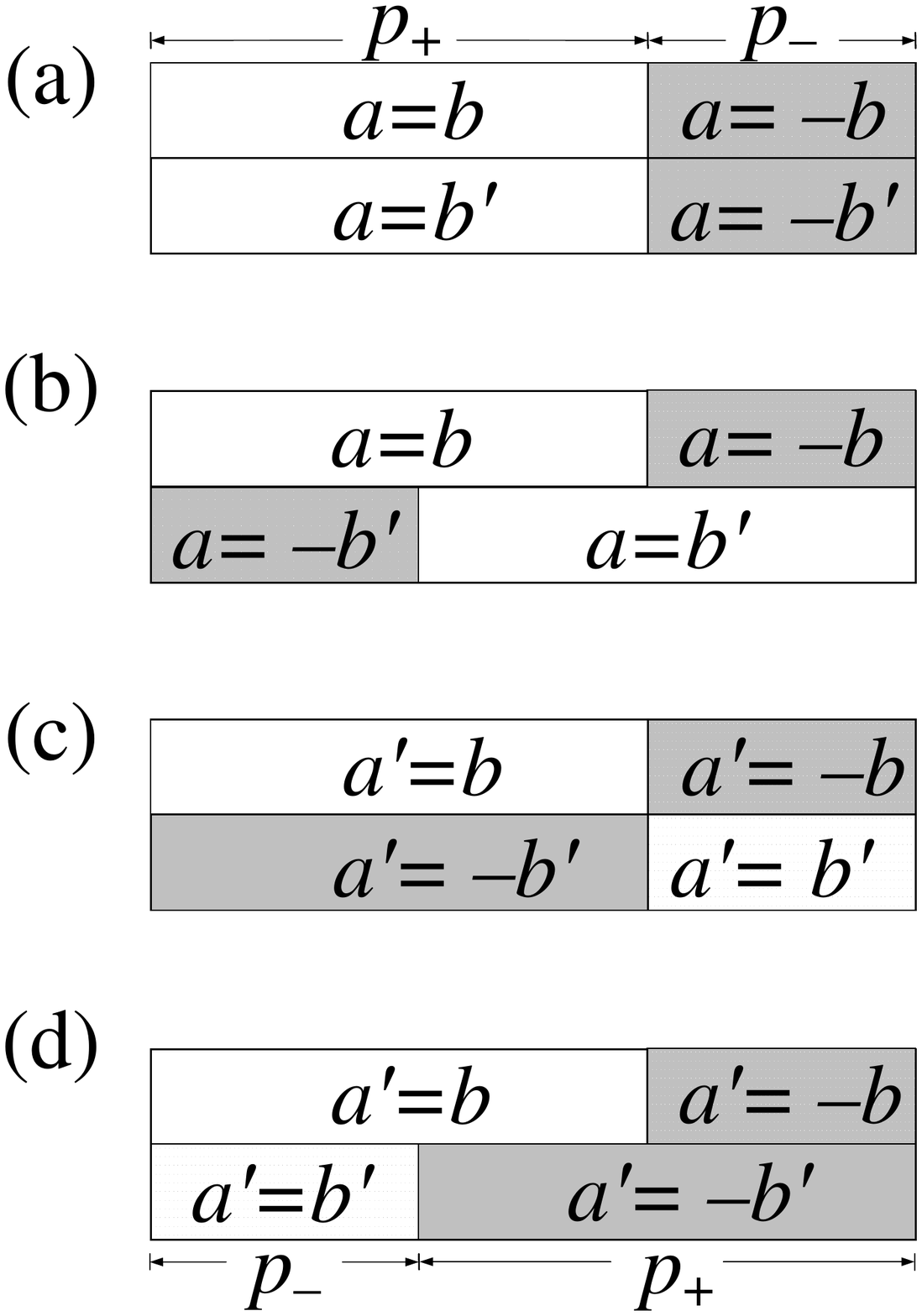}}
\caption[]{(a) $\Delta_a (B+B^\prime)$ maximal;  (b) $\Delta_a (B+B^\prime)$ minimal;
(c) $\Delta_{a^\prime} (B+B^\prime)$ minimal;  (d) $\Delta_{a^\prime} (B+B^\prime)$ maximal.}
\label{Fig1}
\end{figure}


\begin{references}
\bibitem{von}J. von Neumann, {\it Mathematical Foundations of Quantum Mechanics}, trans. R. T
Beyer (Princeton:  Princeton U. Press), 1955.

\bibitem{a} Y. Aharonov, unpublished lecture notes.  See also Y. Aharonov and D. Rohrlich,
{\it Quantum Paradoxes: Quantum Theory for the Perplexed} (Weinheim: Wiley-VCH), 2005, Chaps.
6 and 18.

\bibitem{nleq}Y. Aharonov, H. Pendleton and A. Petersen, {\it Int. J. of Theor. Phys.} {\bf
2}, 213 (1969); Y. Aharonov, in {\it Proc. of the Int. Symp. on the Foundations of Quantum
Mechanics}, Tokyo, 1983, p. 10.  See also Y. Aharonov and D. Rohrlich, {\it op. cit.}, Chaps.
5, 6 and 13.

\bibitem{bell} J. S. Bell, {\it Physics} {\bf 1}, 195 (1964); J. F. Clauser, M. A. Horne, A.
Shimony, and R. A. Holt, {\it Phys. Rev. Lett.} {\bf 23}, 880 (1969).

\bibitem{s}A. Shimony, in {\it Foundations of Quantum Mechanics in Light of the New
Technology}, S. Kamefuchi et al. eds. (Tokyo:  Japan Physical Society), 1983, p. 225; A.
Shimony, in {\it Quantum Concepts of Space and Time}, R. Penrose and C. Isham, eds. (Oxford:
Clarendon Press), 1986, p. 182.

\bibitem{PR} S. Popescu and D. Rohrlich, {\it Found. Phys.} {\bf 24}, 379 (1994).  See also
D. Rohrlich, in {\it Probability in Physics}, eds. Y. Ben-Menahem and M. Hemmo (Berlin: Springer), 2012, pp. 187-200.

\bibitem{o} W. van Dam, {\it  Nonlocality \& Communication Complexity} (Ph.D. thesis), Oxford
University (2000); preprint quant-ph/0501159 (2005); D. Dieks, {\it Phys. Rev.} {\bf A66}, 062104 (2002); H. Buhrman and S. Massar, {\it Phys. Rev.} {\bf A72}, 052103 (2005);  J. Barrett and S. Pironio, {\it Phys. Rev. Lett.} {\bf 95}, 140401 (2005); G. Brassard, H. Buhrman, N. Linden, A. A. M{\'e}thot, A. Tapp and F. Unger, {\it Phys. Rev. Lett.} {\bf 96}, 250401 (2006); J. Barrett, {\it Phys. Rev.} {\bf A75}, 032304 (2007); D. Gross, M. M{\"u}ller, R. Colbeck and O. C. O. Dahlsten, {\it Phys. Rev. Lett.} {\bf 104}, 080402 (2010).

\bibitem{ic}M. Paw{\l}owski et al., {\it Nature} {\bf 461}, 1101 (2009).

\bibitem{prncl} D. Rohrlich, in {\it Quantum Theory: A Two-Time Success Story} (Yakir Aharonov Festschrift), eds. D. C. Struppa and J. M. Tollaksen (New York: Springer), 2013, pp. 205-211

\bibitem{nw}M. Navascu{\'e}s and H. Wunderlich, {\it Proc. R. Soc. A} {\bf 466}, 881 (2010).

\bibitem{bs}N. Brunner and P. Skrzypczyk,  {\it Phys. Rev. Lett.} {\bf 102}, 160403 (2009).

\bibitem{ts}B. S. Tsirelson (Cirel'son), {\it Lett. Math. Phys.} {\bf 4}, 93 (1980).

\bibitem{EPRB}A. Einstein, B. Podolsky and N. Rosen, {\it Phys. Rev.} {\bf 47}, 777 (1935);
N. Bohr, {\it Phys. Rev.} {\bf 48}, 696 (1935).

\bibitem{weak}Y. Aharonov, D. Z. Albert, and L. Vaidman, {\it Phys. Rev. Lett.} {\bf 60},
1351 (1988); see also Y. Aharonov and D. Rohrlich, {\it op. cit.}, Chaps. 16-17.

\bibitem{kb}I thank Konrad Banaszek for suggesting this shortcut.

\end{references}
\end{document}